\documentclass{aa} %[referee]

\usepackage{xcolor}
\usepackage[colorlinks]{hyperref}
\hypersetup{ colorlinks, linkcolor=blue, citecolor=blue }

\hypersetup{
        colorlinks=true,
        allcolors=blue,
        }
\usepackage{natbib}
\usepackage[varg]{txfonts}
\usepackage{graphicx} %use graph format
\usepackage{ulem}

\begin{document}

\title{Irradiation effect and binary radio pulsars with giant companions}

\titlerunning{Irradiation effect and binary radio pulsars with giant companions}
\authorrunning{Lan \& Meng}

\author{Shunyi Lan \inst{1,2}
\and Xiangcun Meng\inst{1,3}
}
\institute{Yunnan Observatories, Chinese Academy of Sciences, Kunming 650011, PR China\\e-mail: lanshunyi@ynao.ac.cn, xiangcunmeng@ynao.ac.cn
\and University of the Chinese Academy of Sciences, Beijing 100049, PR China
\and International Centre of Supernovae, Yunnan Key Laboratory, Kunming 650216, PR China
}

%%\date{5 July 2020}

\abstract
 {Pulsars are neutron stars that rotate rapidly. Most of the pulsars in binary systems tend to spin faster than those in isolation. According to binary evolution theory, radio pulsars in binary systems can have various types of companion stars. However, binary radio pulsars with giant stars, helium stars, and black holes as companions are lacking so far.}
 {We aim to investigate the possible parameter space of binary radio pulsars with giant companions.}
 {We used the MESA stellar evolution code to consider the effect of irradiation in binary evolution and evolved a series of binary models.}
 {We present the potential physical properties of binary radio pulsars with giant star companions in the framework of the classical recycling scenario and the irradiation model. We found that the parameter space of binary radio pulsars with giant companions can be greatly expanded by the effect of irradiation. Moreover, when irradiation is strong enough, submillisecond pulsars might be accompanied by giant companion stars. We also demonstrate a correlation between the timescale of the binary system as a low-mass X-ray binary and the irradiation efficiency. }
 {The birthrate problem between millisecond pulsars and low-mass X-ray binaries might not be resolved by the irradiation effect. A more detailed binary population synthesis is necessary. Our findings may offer guidance for observers that wish to locate binary radio pulsars with giant companions or submillisecond pulsars.}

\maketitle

\section{Introduction}
\label{sec1}

A pulsar is a type of neutron star (NS) that spins rapidly. The first pulsar was discovered in 1967 \citep{1968Natur.217..709H}. Since then, over 3,300 pulsars have been discovered, and the count continuously increases \citep{2005AJ....129.1993M}. Most millisecond pulsars (MSPs), which have a spin period $P_{\rm spin}$ shorter than 30 milliseconds, are found in binary systems. It is well established that these MSPs have increased their spin rate by accreting mass from their low-mass companion star, causing them to become recycled pulsars. During the mass-transfer phase, these systems can be observed as low-mass X-ray binaries (LMXBs). The discoveries of transitional millisecond pulsars and accreting millisecond X-ray pulsars (AMXPs) have strongly supported this scenario \citep[e.g.,][]{2010MNRAS.407.2575P,2013Natur.501..517P}.

The general picture of binary evolution is well established. Starting from a binary of two zero-age main-sequence (ZAMS) stars, the initially more massive star first dies in a supernova event and leaves an NS+MS or black hole (BH) + MS binary system if the binary system remains bound \citep[e.g.,][]{1995MNRAS.274..461B, 2019A&A...624A..66R}. When the orbital period is short enough, mass transfer and mass loss can occur. The form of the mass transfer varies depending on the orbital period and evolutionary stage of the companion stars. For a companion mass $M_2>10~M_\odot$, the mass transfer is dominated by the stellar wind. These systems can be observed as high-mass X-ray binaries (HMXBs), and they may be the progenitors of double neutron stars (DNSs) or NS+BH \citep[e.g.,][]{2017ApJ...846..170T}. When the mass-transfer process of NS+MS or BH+MS is dynamically unstable, the envelope of the companion star may swallow the NS or BH and enter the so-called common-envelope (CE) phase \citep{2013A&ARv..21...59I}. The CE can be ejected under certain conditions and leave an NS+helium star system or a BH+helium star system. For $M_2<10~M_\odot$, the mass transfer is normally dominated by Roche-lobe overflow (RLO), and these systems can be observed as intermediate-mass X-ray binaries (IMXBs) or LMXBs. NS IMXBs lead to the formation of MSPs with carbon-oxygen WD (CO WD) or oxygen-neon-magnesium WD (ONeMg WD)  companions, and NS LMXBs lead to the formation of MSPs with helium WD (He WD) companions \citep[e.g.,][]{1999A&A...350..928T,2012MNRAS.425.1601T}. For other formation channels and a more detailed binary evolution review, we refer to \cite{2020RAA....20..161H} and \cite{tauris2023physics}.

The main framework has been established, but certain physical details still need to be addressed. \cite{1991Natur.350..136P} found that the evolution of LMXBs can be significantly altered by the irradiation effect. \cite{1994ApJ...434..283H} showed that when the irradiation effect is considered, episodic mass transfer occurs. Further research on the detailed mechanism of the irradiation effect was conducted by \cite{2000A&A...360..969R} and \cite{2004A&A...423..281B}, revealing that the irradiation effect is vital not only for LMXBs, but also for accreting WD binaries \citep[e.g.,][]{2022A&A...666A..81Z} . By applying this effect, a series of works by \cite{2012ApJ...753L..33B,2015ApJ...798...44B,2014ApJ...786L...7B,2017A&A...598A..35B} showed that the irradiation effect could play a critical role in the formation of black widow and redback pulsars. Black widows might be evolving from redbacks. The irradiation effect may also account for the formation of AMXPs, as stated in \cite{2020MNRAS.495..796G}.

The irradiation effect on a binary system is strong because it causes mass transfer to occur in a cyclic manner \citep[e.g.,][]{2014ApJ...786L...7B}. This enables the system to accommodate a radio pulsar before the mass transfer ceases completely. According to theory, binary radio pulsars can have companions of various types, such as MSs, giant stars, helium stars, He WDs, CO WDs, ONeMg WDs, NSs, and BHs. However, binary pulsars with companions that are giant stars, helium stars, and BHs are yet to be discovered \citep{2005AJ....129.1993M}. This paper aims to provide the possible parameter space for binary radio pulsars with giant companion stars, with and without the irradiation effect.

In Sect.~\ref{sec2} we introduce our methods and the input physics for numerical calculations. In Sect.~\ref{sec3} we present our results. The discussions are given in Sect.~\ref{sec4}, and we conclude in Sect.~\ref{sec5}.

\section{Input physics and methods}
\label{sec2}

We used the stellar evolution code Modules for Experiments in Stellar Astrophysics \citep[\textsc{MESA}, version 10398, ][]{2011ApJS..192....3P,2013ApJS..208....4P,2015ApJS..220...15P,2018ApJS..234...34P,2019ApJS..243...10P} to run evolutions of LMXBs. In all of our calculations, we assumed an initial NS mass of $1.4~M_\odot$. The companion stars were all ZAMS stars, and the masses ranged from $1.0$ to $2.0~M_\odot$ with a step of $0.1~M_\odot$. The metallicity of the companion star was $Z=0.02$. We considered the prescription of \cite{1975MSRSL...8..369R} for the stellar wind, and we set the scaling factor to 0.1. The initial orbital periods of binary systems, $\log P_{\rm orb}/{\rm days}$, range from $0.5$ to $2.0$. We adopted the \cite{1990A&A...236..385K} scheme to calculate the mass-transfer rate of RLO. We adopted the isotropic reemission model \citep{tauris2023physics} to compute the mass loss from the binary system with the following parameters: $\alpha=0,~\beta=0.7, and ~\delta=0$, which correspond to the fraction of the mass that is lost in the vicinity of the companion star, in the vicinity of the NS, and in the circumbinary coplane toroid, respectively. The mass-accretion rate onto the NS is $\dot{M}_{\rm NS}=(1-\alpha-\beta-\delta)|\dot{M_2}|$, where $\dot{M}_2$ is the mass-loss rate of the companion star. We did not consider mass accretion of the stellar wind of the companion star, and $\dot{M}_{\rm NS}$ was limited by the Eddington accretion rate \citep{2012MNRAS.425.1601T},
    \begin{equation}
        \dot{M}_{\rm Edd} \simeq3.0\times10^{-8}M_\odot\mathrm{yr}^{-1}~ \left(\frac{R_{\rm NS}}{13~{\rm km}} \right)\left(\frac{1.3}{1+X}\right),
    \end{equation}
where the $R_{\rm NS}=15~(M_{\rm NS}/M_\odot)^{1/3}~{\rm km}$ is the radius of the NS, and $X$ is the fraction of hydrogen in the transferred materials.

The spin evolution of the NS was implanted and calculated self-consistently with the MESA code. We calculated the magnetospheric radius $r_{\rm mag}$, the corotation radius $r_{\rm co}$, and the light cylinder radius $r_{\rm lc}$ as \citep{frank2002accretion}
\begin{equation}r_{\rm mag}=\xi\left(\frac{\mu^{4}}{2GM_{\rm NS}\dot{M}_{\rm NS}^{2}}\right)^{1/7},\end{equation}

\begin{equation}
    r_{\rm co} = \left(\frac{GM_{\rm NS}P_{\rm spin}^2}{4\pi^2} \right)^{\frac{1}{3}},
\end{equation}

\begin{equation}
    r_{\rm lc} = \frac{cP_{\rm spin}}{2\pi},
\end{equation}
where $\xi$ is a dimensionless constant depending on the details of the disk-magnetosphere interactions, which is unknown for now. We took $\xi=0.5$ in this paper. $G$ is the gravitational constant, and $\mu$ is the magnetic moment of the NS. $M_{\rm NS}$ and $\dot{M}_{\rm NS}$ are the mass and accretion rate of the NS, respectively. $P_{\rm spin}$ is the spin period of the NS, and $c$ is the speed of light in vacuum.

We implemented the irradiation effect by injecting additional energy into the outer envelope of the companion star. The amount of this energy was determined by the irradiation luminosity $L_{\rm irr}$, and we followed \cite{1994ApJ...434..283H},
    \begin{equation}
        % \begin{aligned}
 \\
L_{\rm irr} =\left\{\begin{array}{l l}{{\eta L_{\rm X}\left(\frac{R_{2}}{2a}\right)^{2}}}&{{\dot{M}<\dot{M}_{\rm Edd}}}\\ {{\eta L_{\rm X}\left(\frac{R_{2}}{2a}\right)^{2}\exp\left(1-\frac{|\dot{M}|}{\dot{M}_{\rm Edd}}\right)}}&{{\dot{M}\geq\dot{M}_{\rm Edd}}}\end{array}\right., 
        % \end{aligned}
    \end{equation}
where $\eta$ is the irradiation efficiency, $L_{\rm X}$ is the accretion luminosity, $R_2$ is the radius of the companion star, $a$ is the binary separation, $\dot{M}$ is the mass-transfer rate, and $\dot{M}_{\rm Edd}$ is the Eddington limit.

The maximum age of the system was set to 14 Gyr. For the other details of physics that are not mentioned here, we followed \cite{2023ApJ...956L..24L}.

\section{Results}
\label{sec3}

\begin{figure*}
        \includegraphics[width=17cm]{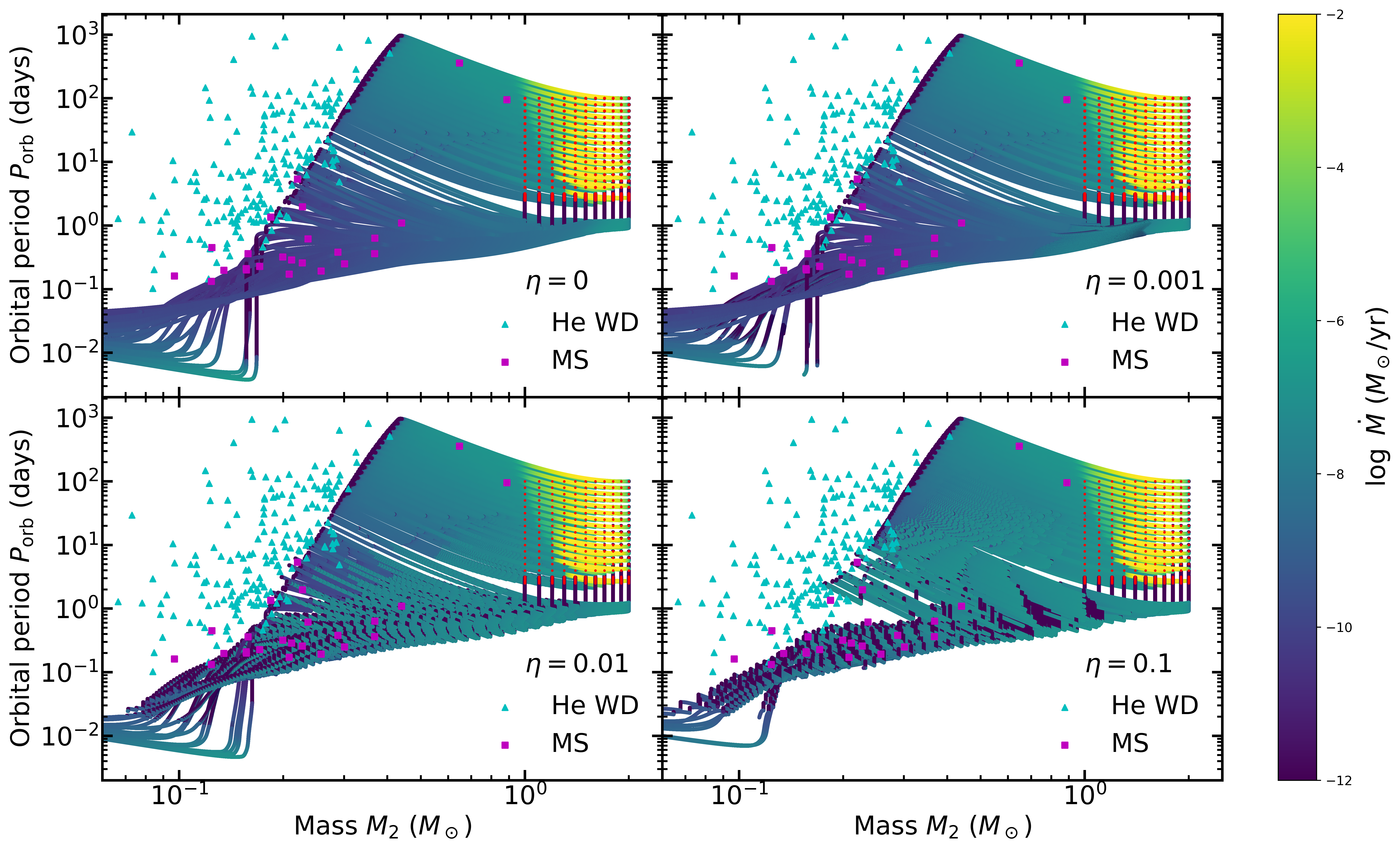}
    \caption{Binary evolution results in a $M_2-P_{\rm orb}$ diagram. The observed binary pulsars with He WDs and MSs are plotted as cyan upward-pointing triangles and magenta squares, and the masses are the lowest possible masses of the companions of observed binary pulsars. The red dots indicate the initial grid. The color bar indicates the mass-transfer rate. The different panels show different irradiation efficiencies.}
    \label{fig1}
\end{figure*}

Figure~\ref{fig1} shows our binary evolution results in a $M_2-P_{\rm orb}$ diagram for $\mu_{\rm i}=1.0\times10^{30}~{\rm G~cm^3}$. The dots that represent observation indicate the minimum mass of the companion star, and we did not plot error bars for clarity. The different panels show different irradiation efficiencies $\eta $ of $ 0,~0.001,~0.01, and ~0.1$. The deep blue segments indicate that the mass-transfer rate is very low, that is, the system is a detached binary. The figure shows that when the irradiation is weak, for example, $\eta=0.001$, the overall evolution is similar to the results of $\eta=0$. For $\eta=0.01 and ~0.1$, the mass-transfer cycles are more evident and the average mass-transfer rate is higher than the results of $\eta=0,~0.001$, as shown by some previous studies \citep[e.g.,][]{2014ApJ...786L...7B,2015ApJ...798...44B}. Some evolutionary tracks for converging systems were stopped manually. We only focused on systems that are able to have giant companions.

\subsection{Irradiation-induced mass-transfer cycles}

\begin{figure*}
        \includegraphics[width=17cm]{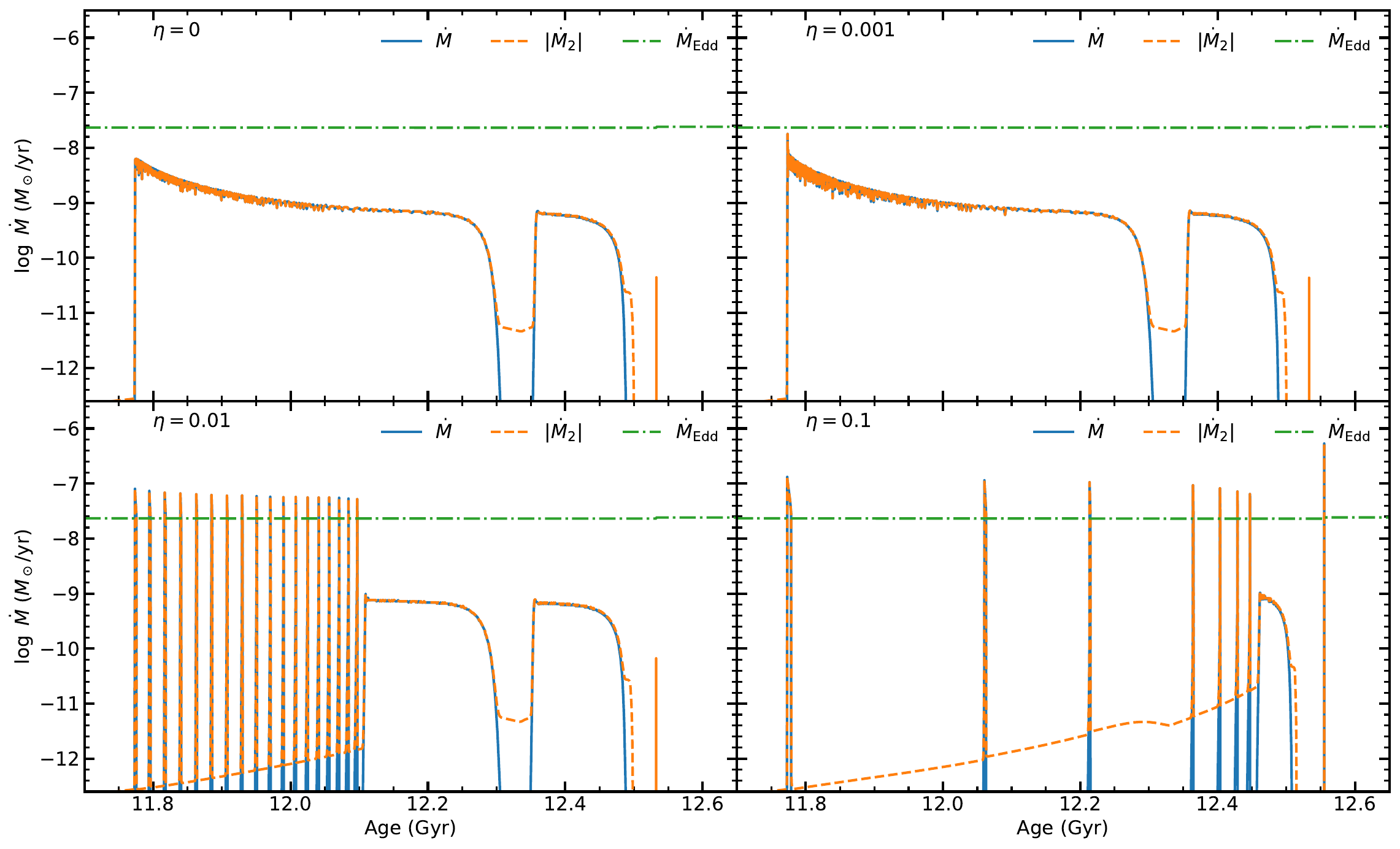}
    \caption{Evolution of the mass-transfer rate for different $\eta$. The initial companion mass is $M_2=1.0~M_\odot$, and the initial orbital period is $\log~P_{\rm orb}/{\rm days}=0.5$. The initial magnetic moment of the NS is $\mu_{\rm i}=1.0\times10^{30}~{\rm G~cm^3}$. The blue, dashed orange, and dash-dotted green lines show the mass-transfer rate, the mass-loss rate of the companion, and the Eddington limit, respectively. }
    \label{fig2}
\end{figure*} 

\begin{figure*}
        \includegraphics[width=17cm]{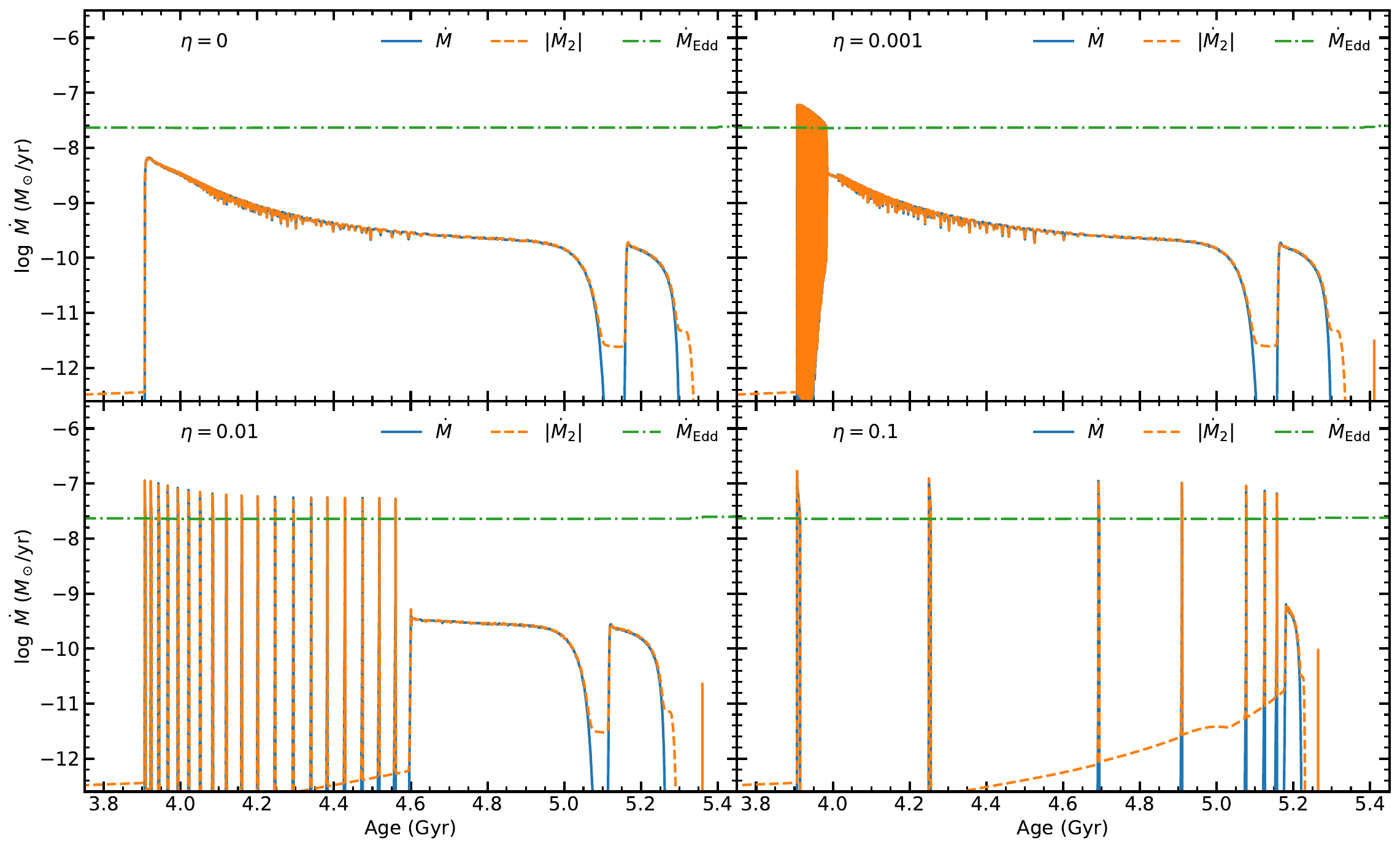}
    \caption{Similar to Fig. \ref{fig2}, but for $(M_2/M_\odot,\log~P_{\rm orb}/{\rm days})=(1.3,0.5)$. Many mass-transfer cycles lie in the orange region for the case of $\eta=0.001$.}
    \label{fig_app_1}
\end{figure*}

\begin{figure*}
        \includegraphics[width=17cm]{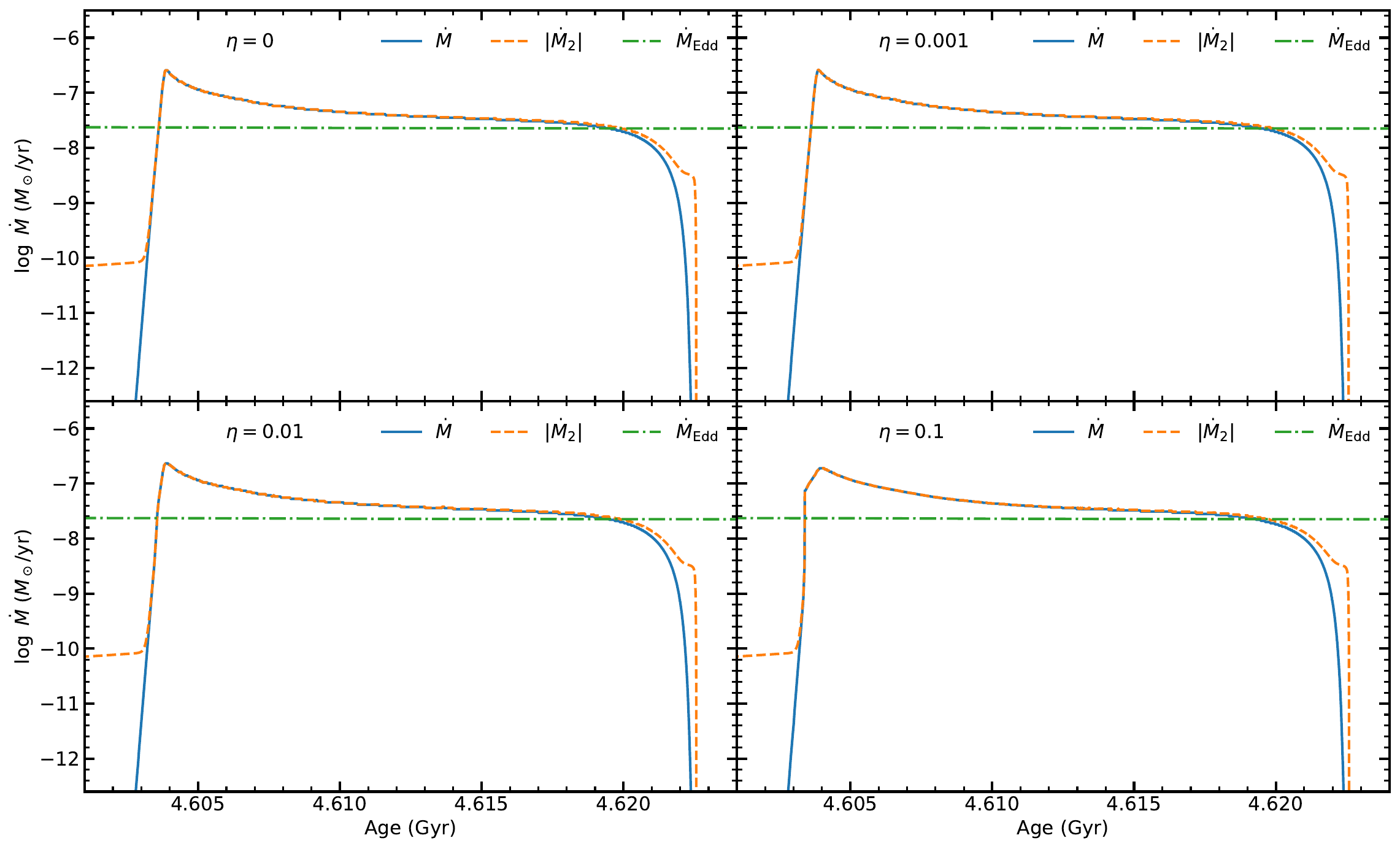}
    \caption{Similar to Fig. \ref{fig2}, but for $(M_2/M_\odot,\log~P_{\rm orb}/{\rm days})=(1.3,1.5)$.}
    \label{fig_app_2}
\end{figure*}

In Fig.~\ref{fig2} we show an example for the evolution of the mass-transfer rate as a function of time for different $\eta$ . The initial companion mass is $M_2=1.0~M_\odot$ and initial orbital period is $\log~P_{\rm orb}/{\rm days}=0.5$. The initial value of the NS magnetic moment is $\mu_{\rm i}=1.0\times10^{30}~{\rm G~cm^3}$. For $\eta=0$, the mass transfer is continuous. The mass transfer was only halted once in the middle, when the hydrogen-burning shell encountered the layer in which the outer convective zone reached its deepest penetration, creating a molecular weight discontinuity. This reduction in molecular weight caused the companion star to contract, and this interrupted the mass-transfer process \citep{2013sse..book.....K}. For the case with weak irradiation, that is, $\eta=0.001$, the irradiation effect on the companion star can be easily offset by thermal relaxation. The mass-transfer rate of $\eta=0.001$ therefore evolves in a similar way to that of the nonirradiated star. For $\eta=0.01$, the thermal equilibrium is disrupted by the stronger irradiation, causing the companion star to expand and the mass-transfer rate to exceed the Eddington limit. When the expansion due to irradiation is outweighed by the thermal relaxation of the companion star, the companion star almost immediately detaches from its RL. The system remains a binary radio pulsar until the companion star refills its RL. For more details about irradiation-induced mass-transfer cycles, we refer to \cite{2000A&A...360..969R}. For $\eta=0.1$, the irradiation is even stronger, and it is quite difficult to reach thermal equilibrium for the companion stars. The mass-transfer rate can reach a higher peak, and the system looses more mass within a single cycle. The reattachment timescale is therefore longer and the number of mass-transfer cycles is smaller. 

We show another two examples of the mass-transfer evolution in Figs. \ref{fig_app_1} and \ref{fig_app_2} to illustrate the impact of the initial companion mass and orbital period on the evolution. In Fig. \ref{fig_app_1}, the initial companion mass is $M_2=1.3~M_\odot$ and the initial orbital period is $\log~P_{\rm orb}/{\rm days}=0.5$. The mass transfer initiates at the Hertzsprung gap (HG), and for this system, even for weak irradiation, that is, $\eta=0.001$, the irradiation-induced mass-transfer cycles may also occur at an early stage of the mass transfer because the companion star has a thicker convective envelope than a $1~M_\odot$ star (see the panel of $\eta=0.001$). The later evolution of the mass-transfer rate is similar to that of $\eta=0$ because the companion star is more evolved and the irradiation can easily be relaxed. For $\eta=0.01$ and 0.1, the evolution of the mass transfer is similar to that of \ref{fig2}. In Fig. \ref{fig_app_2}, the initial companion mass is $M_2=1.3~M_\odot$ and the initial orbital period is $\log~P_{\rm orb}/{\rm days}=1.5$. The irradiation cannot significantly affect a highly evolved giant star as a companion because the thermal timescale of the envelope is very short and the irradiation effect can easily be relaxed by the companion star.

\subsection{Binary radio pulsars with giant companions}

\begin{figure*}
        \includegraphics[width=17cm]{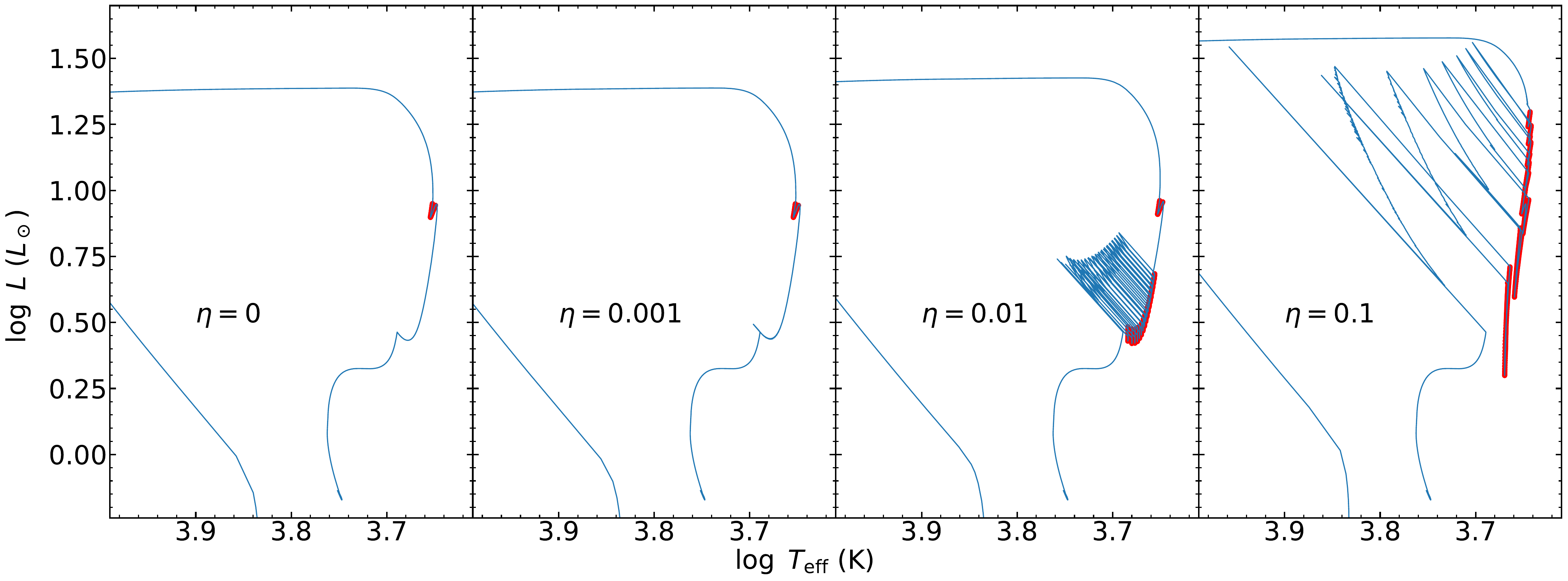}
    \caption{Corresponding evolution of Fig.~\ref{fig2} in the HR diagram. The different panels represent various irradiation efficiencies. The red dots indicate $r_{\rm mag}>r_{\rm lc}$, i.e., the system can be considered as a binary radio pulsar.}
    \label{fig3}
\end{figure*}

\begin{figure}
        \centering
    \resizebox{\hsize}{!}{\includegraphics{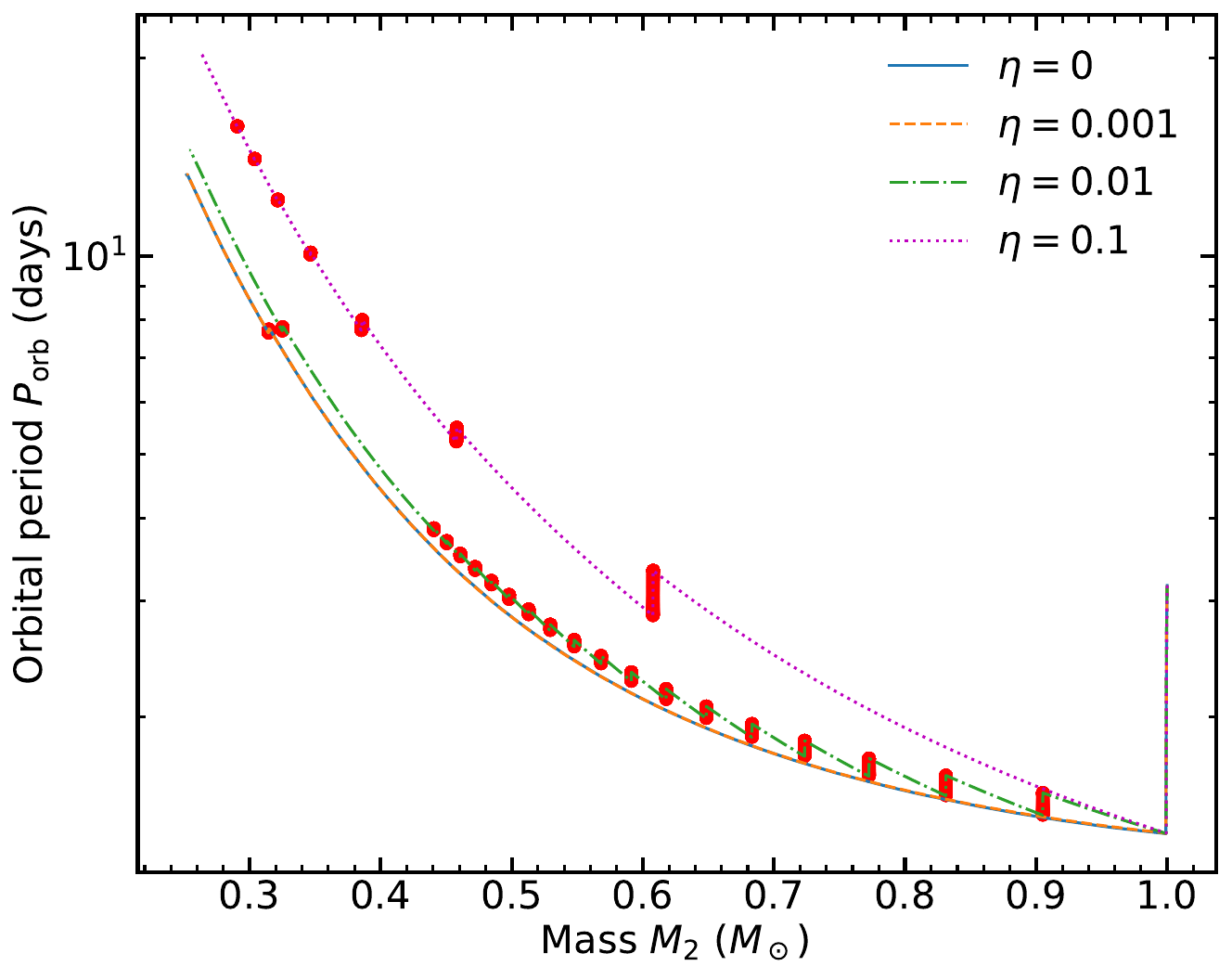}}
    \caption{Corresponding evolution of Fig.~\ref{fig2} in the $M_2-P_{\rm orb}$ diagram. The differently colored lines indicate different irradiation efficiencies. The red dots indicate $r_{\rm mag}>r_{\rm lc}$, i.e., the system can be considered as a binary radio pulsar.}
    \label{fig4}
\end{figure}

In order to intuitively present the evolutionary state of a binary radio pulsar with a giant companion star, we show in Figs.~\ref{fig3} and \ref{fig4} the corresponding evolution of the system in Fig.~\ref{fig2} in a Hertzsprung-Russell (HR) diagram and a $M_2-P_{\rm orb}$ diagram, respectively. In Fig.~\ref{fig3}, the cases with different $\eta$ are shown in different panels, while in Fig.~\ref{fig4}, differently colored lines correspond to distinct irradiation efficiencies. In both figure the red dots indicate $r_{\rm mag}>r_{\rm lc}$, i.e., the system can be considered as a binary radio pulsar. Compared to the evolution of weak irradiated $\eta=0.001$ or non-irradiated model, for $\eta=0.01,~0.1$, when mass transfer turns on, the position of a companion star in the HR diagram deviates from red giant branch (RGB) to become brighter and hotter. The additional energy provided by irradiation that is stored on the surface can travel inward or radiate off the surface of the companion star and causes the stellar luminosity to rise. As the irradiation pushes the companion star out of thermal equilibrium, it expands, which causes its effective temperature to rise at the same time. The stronger the irradiation, the larger the deviation from the RGB for the companion star. In Fig.~\ref{fig4} we show for a given evolutionary stage that the orbital period of a binary system that is strongly irradiated is longer than that of a system that is weakly irradiated or not irradiated at all. The reason is that when the irradiation is strong, the mass-transfer rate can exceed the Eddington limit, which results in the ejection of more mass from the binary system, as shown in Fig.~\ref{fig2}. Moreover, the final mass of the WD is slightly higher because the final orbital period is longer because the system still follows the relation between the mass of the WD and the binary orbital period \citep[e.g.,][]{1999A&A...350..928T}.

\begin{figure*}
        \centering
    \includegraphics[width=17cm]{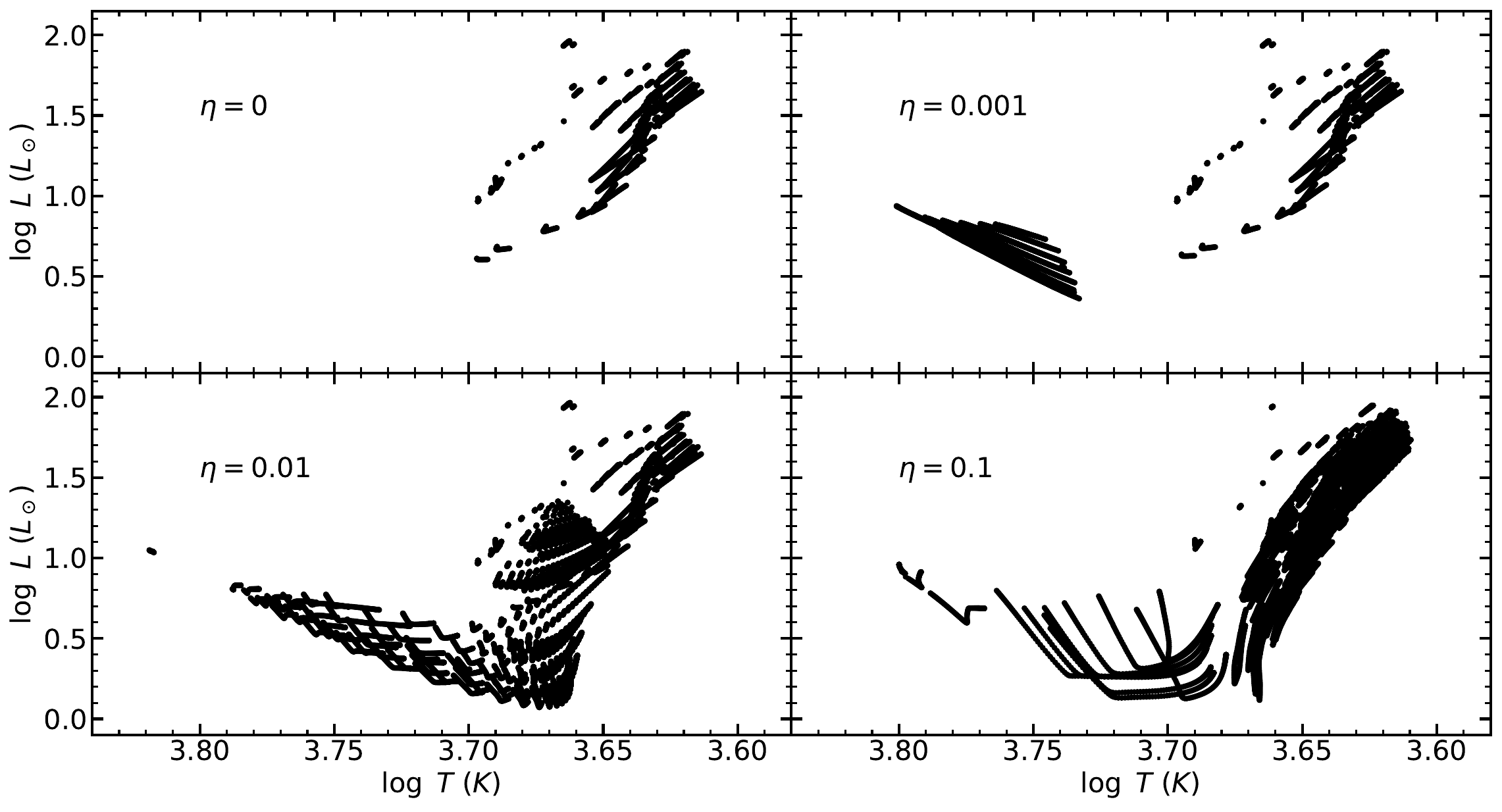}
    \caption{HR diagram showing the parameter space for giant stars, which serve as the companions of radio pulsars. The different panels show different $\eta$.}
    \label{fig5}
\end{figure*}

\begin{figure*}
        \centering
    \includegraphics[width=17cm]{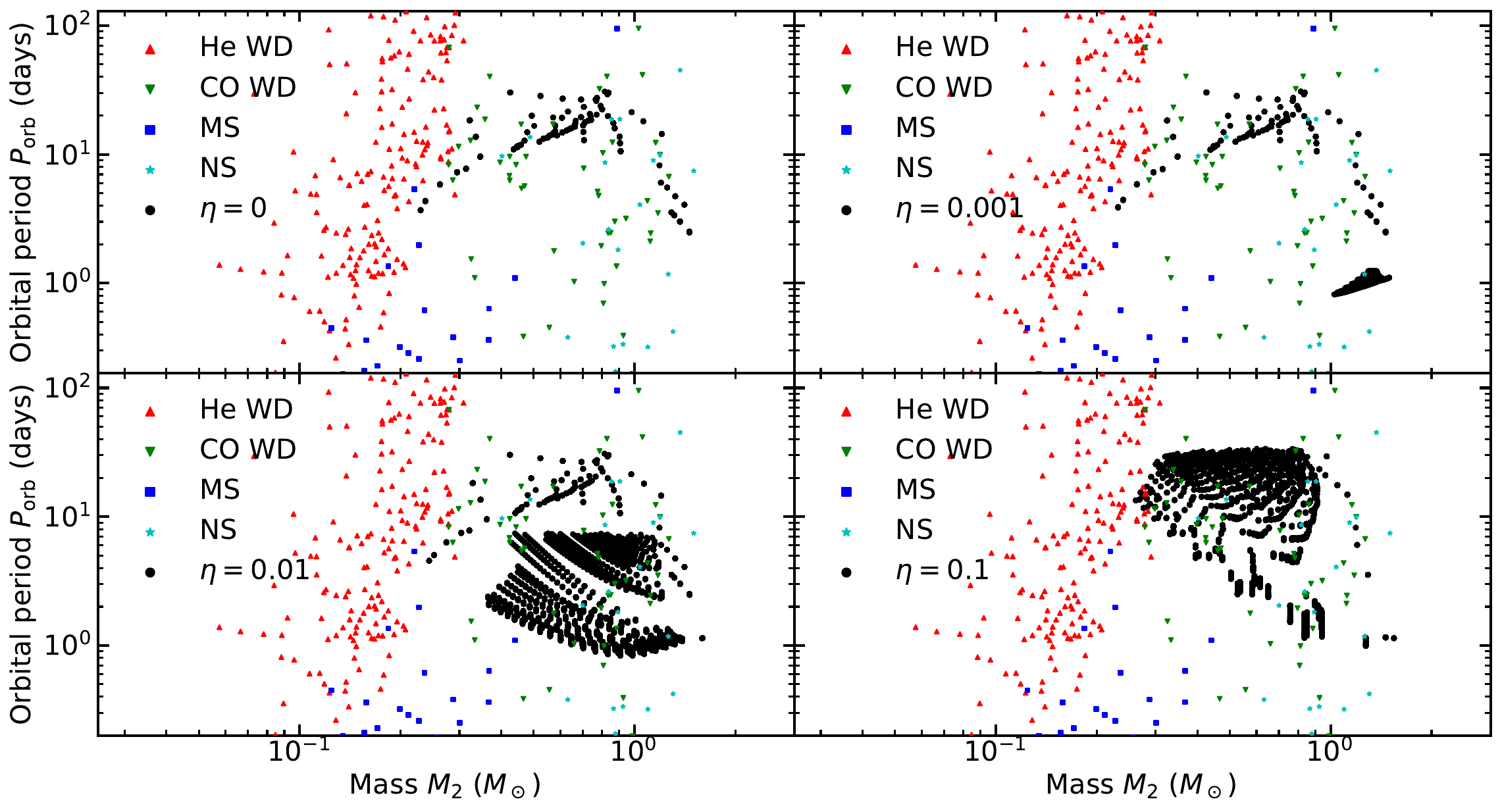}
    \caption{Possible parameter space of binary radio pulsars in a $M_2-P_{\rm orb}$ diagram for different $\eta$. The upperward-pointing red and downward-pointin green triangles, blue squares, and cyan stars indicate observed binary pulsars with He WDs, CO WDs, MSs, and NSs. The data are taken from the ANTF pulsar catalog \citep{2005AJ....129.1993M}. The black dots are the results we calculated for binary radio pulsars with giant companions.}
    \label{fig6}
\end{figure*}

\begin{figure*}
        \centering
    \includegraphics[width=\textwidth]{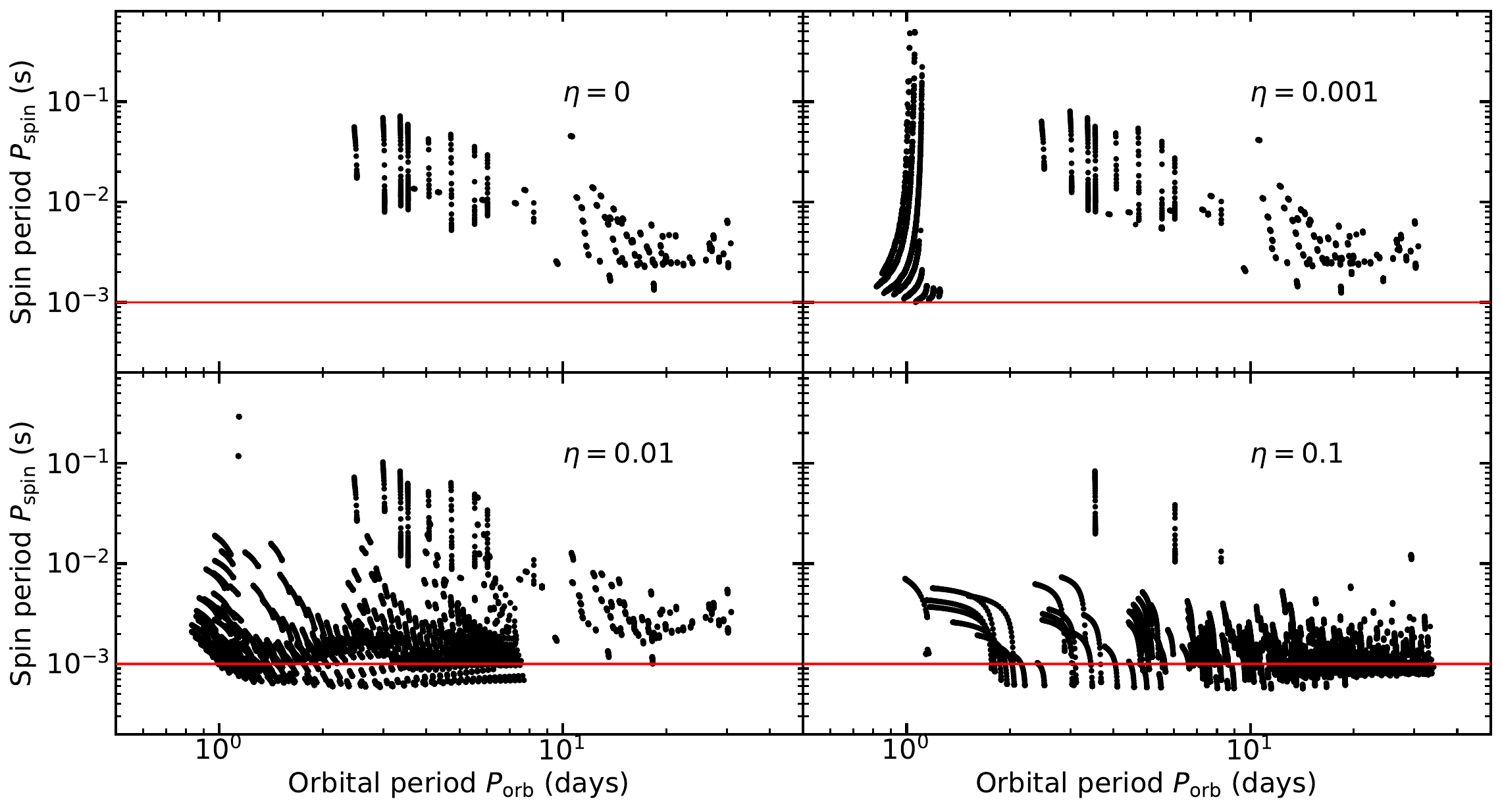}
    \caption{Possible parameter space of binary radio pulsars with giant companions in a Corbet diagram. The different panels show different $\eta$. The horizontal red line indicates a spin period of 1 ms.}
    \label{fig7}
\end{figure*}

Figure~\ref{fig5} shows the possible parameter space of giant star companions of binary radio pulsars in the HR diagram for different $\eta$ . As expected, the parameter space of binary radio pulsars with giant companion stars is expanded through the irradiation effect, that is, the parameter space increases with $\eta$. In our models, the mass transfer can be initiated by a companion star in the HG or on the RGB, which means that the companion stars can be a subgiant or a red giant. In the case of $\eta=0.001$, it is clear that the parameter space is divided into two parts. In Fig.~\ref{fig6}, we show the possible parameter space of binary radio pulsars in a $M_2-P_{\rm orb}$ diagram for different $\eta$ . Observed binary pulsars with different types of companion are also shown in the figure. The parameter space is smallest for $\eta=0$. For $\eta=0.001$, part of its dots coincide with that of $\eta=0$ because most of the evolution with weak irradiation is basically not affected. Some other systems initiate mass transfer on the subgiant branch, and the evolution can be significantly altered by even weak irradiation in the early stage of mass transfer (see Fig. \ref{fig_app_1}). For $\eta=0.01$ and $\eta=0.1$, the parameter space is clearly enlarged due to irradiation-induced mass-transfer cycles.

In Fig. \ref{fig7}, we show the possible parameter space of binary radio pulsars with giant companions in a Corbet diagram, that is, a $P_{\rm orb}-P_{\rm spin}$ diagram. Similar to Figs.~\ref{fig5} and \ref{fig6}, the different panels correspond to different $\eta$. The red horizontal line indicates the spin period of 1 ms. Our results show that when irradiation is strong enough, the pulsar can be accelerated to the submillisecond level. Although our model is simplified (see Sect.~\ref{sec4} for a more detailed discussion), these results imply that the sub-MSPs might be accompanied by giant stars.

% \subsection{Sub-thermal timescale mass transfer and radio MSPs with high spin period derivative}

\subsection{Mass-transfer timescale}

\begin{figure}
        \centering
    \resizebox{\hsize}{!}{\includegraphics{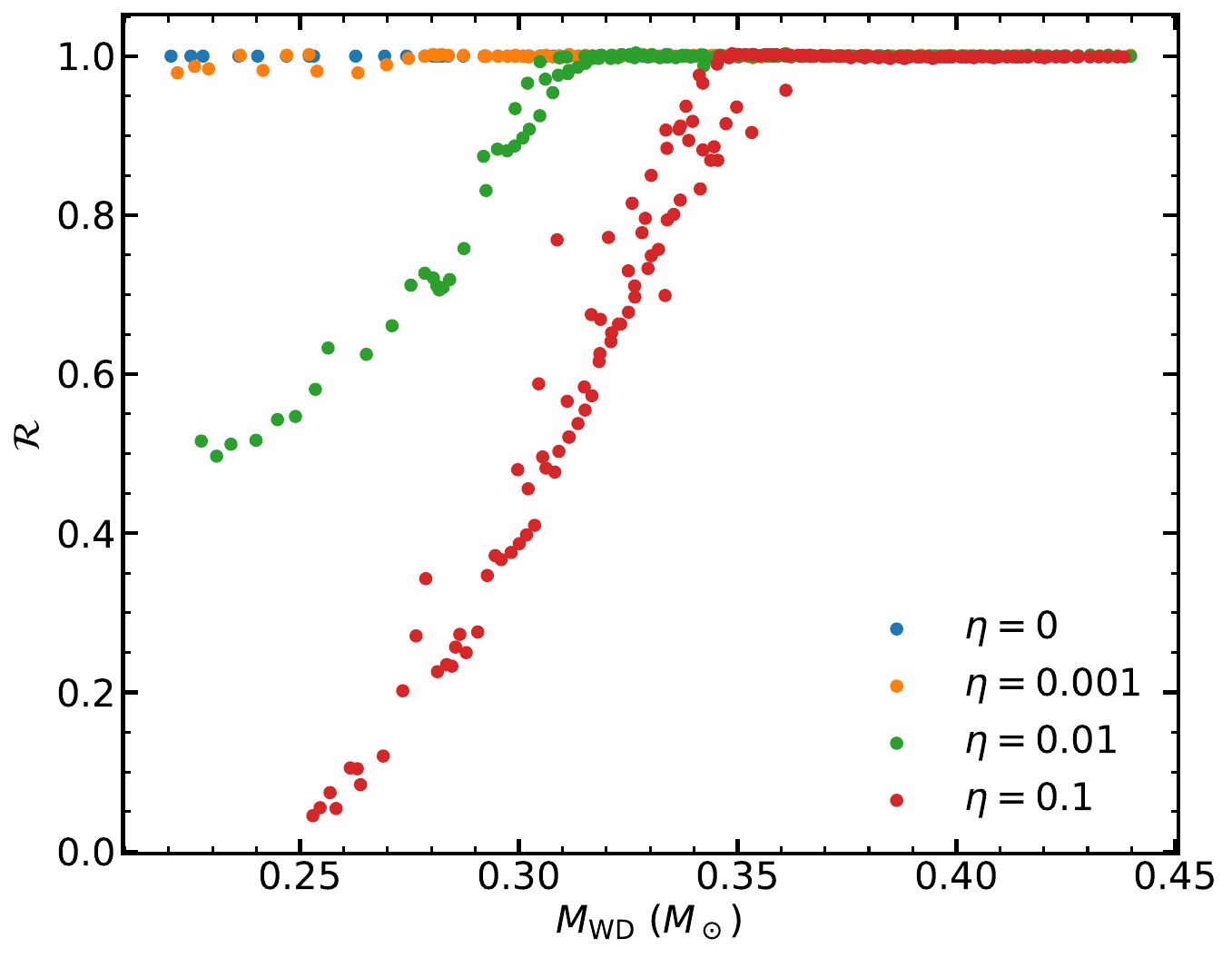}}
    \caption{Relation between $\mathcal{R}$ and the final mass of a WD $M_{\rm WD}$. The dot shows our calculated results, and different colors represent different irradiation efficiencies. See text for the definition of $\mathcal{R}$.}
    \label{fig8}
\end{figure}

\cite{1988ApJ...335..755K} first reported that the birth rates of MSPs and LMXBs do not match. There are two ways to resolve this birthrate problem: The number of LMXBs can be increased through observations, or the system timescale can be reduced for an LMXB in theory. In the context of the classical recycling scenario, that is, $\eta=0$, the system timescale for LMXBs, $\tau_{\rm LMXB}$, typically ranges from $10^7$ to $10^{10}$ yr, which depends on the initial parameters of a binary \citep[e.g., mass of companion or orbital period, see also][]{2002ApJ...565.1107P}.

As previously demonstrated, the irradiation effect can significantly reduce $\tau_{\rm LMXB}$, as shown in Figs. \ref{fig2}, \ref{fig_app_1}, and \ref{fig_app_2}. The system timescale of LMXBs after considering the irradiation effect, $\tau_{\rm LMXB, irrad}$, not only depends on the initial parameters of the binary (mass of companion and orbital period), but also on the efficiency of the irradiation. We define the parameter $\mathcal{R}$ to characterize the degree to which $\tau_{\rm LMXB}$ is reduced, 
\begin{equation}
    \mathcal{R} = \frac{\tau_{\rm LMXB, irrad}}{\tau_{\rm LMXB}}.
\end{equation}

Since the initial companion mass and orbital period may be related to the final mass of the WD $M_{\rm WD}$ \citep[e.g.,][]{1999A&A...350..928T}, we relate $\mathcal{R}$ to $M_{\rm WD}$ in Fig.~\ref{fig8}. For $\eta=0$, $\mathcal{R}$ is always 1, which corresponds to the canonical recycling scenario. For weak irradiation, $\eta=0.001$, most systems are not affected and $\mathcal{R}$ is very close to 1. For stronger irradiation, $\mathcal{R}$ depends on $M_{\rm WD}$. For a given $\eta$, there is a threshold of $M_{\rm WD}$, that is, when $M_{\rm WD}$ exceeds the threshold, $\mathcal{R}$ is equal to 1, and $\mathcal{R}$ increases with $M_{\rm WD}$ when $M_{\rm WD}$ is lower than the threshold. For $\eta=0.01$, the threshold is $\sim0.31~M_\odot$, while for $\eta=0.1$, it is $\sim0.34~M_\odot$. The smallest $\mathcal{R}$ can be $\sim 0.05$. The reason for these results is that for a given companion star, the irradiation effect  can only be significant when the companion star is in an earlier evolution state, which depends on $\eta$. For a more evolved companion star, the thermal timescale of the envelope is very short, and the irradiation effect can easily be relaxed by the companion star (see Fig. \ref{fig_app_2}).

\section{Discussion}
\label{sec4}
% \subsection{Spin evolution before mass transfer}
% \subsection{Spin-down by stellar wind}

\begin{figure}
        \centering
    \resizebox{\hsize}{!}{\includegraphics{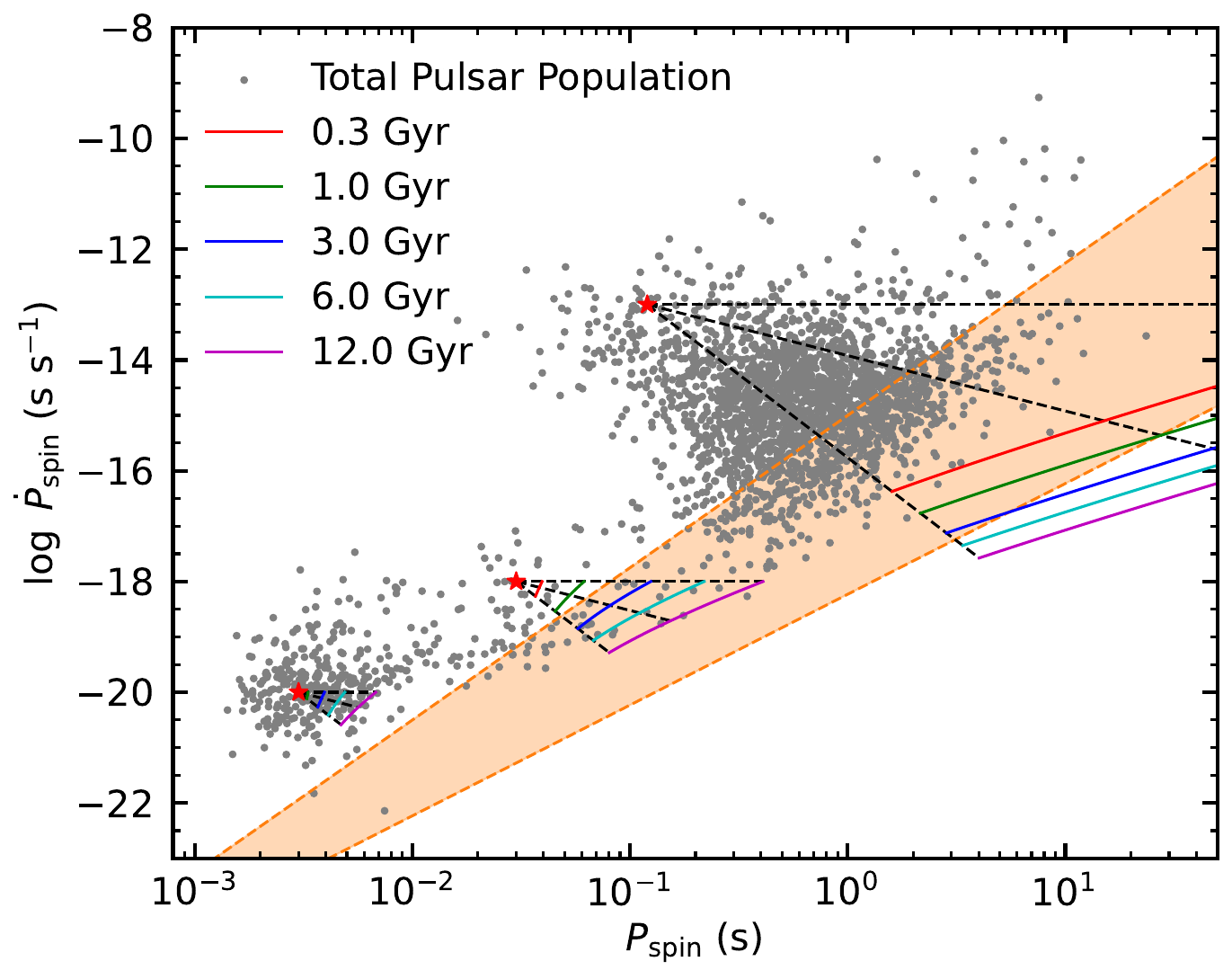}}
    \caption{Isochrones of three hypothetical pulsars. The gray dots show the currently observed population of pulsars. The data are taken from the ANTF pulsar catalog \citep{2005AJ....129.1993M}. The red stars indicate the birth positions of the hypothetical pulsars. The colored lines indicate isochrones and were calculated for different values of the braking index $2\leq n \leq5$. The dashed black lines from top to bottom indicate the evolution of $n=2,~3$, and 5. The shaded orange area indicates the so-called death valley \citep{2024ApJ...961..214R}.}
    \label{fig9}
\end{figure}

We calculated a series of binary evolution models to show the possible parameter space of binary radio pulsars with giant companions. The results we show in Figs.~\ref{fig5}, \ref{fig6}, and \ref{fig7} ignore the evolution before the first RLO, even though the conditions $r_{\rm mag}>r_{\rm lc}$ and the presence of a giant companion star are both met. In our models, binary evolution starts with NS+ZAMS. The range of time needed to initiate the first RLO is $\sim 0.9-12~{\rm Gyr}$. The lifetime of an MS star is a function of mass. For a $2.0~M_\odot$ ZAMS star, it is $\sim 1.0~{\rm Gyr}$, and for a $1.0~M_\odot$ ZAMS star, it is $\sim 10~{\rm Gyr}$ \citep{2013sse..book.....K}. Following \cite{2012MNRAS.425.1601T}, we analyzed the spin evolution of an NS before the first RLO. The isochrones of three hypothetical pulsars are plotted in Fig.~\ref{fig9}. We assumed that the red stars are the birth positions of pulsars in the $P_{\rm spin}-\dot{P}_{\rm spin}$ diagram. The radio emission is quenched when pulsars are within or below a death valley. For the top hypothetical pulsar, it takes less than 0.3 Gyr to reach the death valley, and for the middle hypothetical pulsar, this is $\sim2$ Gyr. For MSPs, the corresponding timescale is longer than 12 Gyr. Although the birth state of an NS remains unclear, if the birth state of the NS is similar to the states of normal single pulsars, the part of the parameter space of binary pulsars with giant companions that contribute to the binary evolution before the first RLO can be neglected.

In our simplification, we omitted the accretion of stellar wind onto the NS. Although the stellar wind from a giant companion may approach $10^{-8}~M_\odot/{\rm yr}$, the accretion efficiency onto an NS is relatively low; it is estimated to be about $10^{-4}$ to $10^{-3}$. Therefore, the accretion rate of the NS by the stellar wind of the companion is $\lesssim 10^{-11}~M_\odot/{\rm yr}$. Instead of spinning up, the NS may instead spin down due to the propeller effect. In our numerical calculation, we also assumed a perfect dipole for a radio pulsar, that is, n=3. As mentioned before, there are uncertainties in the spin-down process (see Fig. \ref{fig9}). The parameter space of binary radio pulsars shown in Fig. \ref{fig7} might therefore be slightly shifted upward (or downward). There are still parts of the parameter space crossing the 1 ms line. This indicates that a radio sub-MSP might be accompanied by a giant star if $\eta$ were high enough. If researchers are consistently unable to detect binary radio pulsars with giant companions or binary radio sub-MSP with giant companions, this suggests that the irradiation effect might not a significant factor in the evolution of LMXBs. Furthermore, this implies that this effect might not resolve the birthrate problem. A more detailed binary population synthesis is necessary to determine whether the irradiation effect may resolve the birthrate problem in the future. We have only explored a limited range of companion masses (1.0-2.0$~M_\odot$), and the parameter space we showed might therefore be biased. A more comprehensive range of companion masses should also be investigated in the future.

Theoretical models predict the existence of binary radio pulsars with giant companions, but these objects remain undetected. According to our model, even when the irradiation effect is accounted for, which expands the parameter space, the timescale for the observation of such a system as a binary radio pulsar remains brief, $\lesssim 0.3$ Gyr. This can make these objects difficult to detect. Another reason why binary radio pulsars with giant companions remain undetected may be the viewing geometry. The mass transfer can cause the spin axis of the pulsar to align with the angular momentum vector of the binary system, and thus, the pulsar may not be detected. Although the parameter space becomes larger, the mass-transfer cycles constantly cause the spin axis of the pulsar to align with the angular momentum vector of the binary, which may make it difficult to observe \citep[e.g.,][]{2014MNRAS.439.2033G}.

\section{Conclusions}
\label{sec5}

We studied the irradiation effect and calculated a series of binary models to present the potential parameter space for binary radio pulsars from a purely binary evolutionary perspective. These models were compared with those that evolve without the irradiation effect. The inclusion of irradiation-induced mass-transfer cycles results in an expanded parameter space compared to models without irradiation, as illustrated in Figs. \ref{fig5}, \ref{fig6}, and \ref{fig7}. The parameter space is very limited when the irradiation effect is not considered or when the irradiation is weak. This may be one reason why binary radio pulsars with giant companions are so difficult to find. Our results suggest that radio sub-MSPs might be accompanied by giant stars when the irradiation is indeed strong in the evolution of LMXBs. We also found a relation between the final mass of WDs and the irradiation efficiencies (see Fig. \ref{fig8}).

The discovery of binary radio pulsars with giant companions can significantly constrain theories of binary evolution. Our findings may be verified by future observations or guide observers in the search for these objects.
%   \begin{enumerate}
%       \item The final orbital period and companion mass are similar to those of evolution that not consider irradiation effect. The 
%          properties of the layer.
%   \end{enumerate}

\begin{acknowledgements}
S.L. sincerely thanks Hai-Liang Chen, Zheng-Wei Liu, Hong-Wei Ge and Wei-Tao Zhao for helpful discussions. This work is supported by the National Natural Science Foundation of China (Nos. 12288102 and 12333008) and National Key R\&D Program of China (No. 2021YFA1600403). X.M. acknowledges support from the Yunnan Ten Thousand Talents Plan - Young \& Elite Talents Project, and the CAS `Light of West China' Program. X.M. acknowledges support from International Centre of Supernovae, Yunnan Key Laboratory (No. 202302AN360001), the Yunnan Revitalization Talent Support Program-Science \& Technology Champion Project (NO. 202305AB350003), Yunnan Fundamental Research Projects (Nos. 202401BC070007 and 202201BC070003) and the science research grants from the China Manned Space Project. The authors also acknowledge the ``PHOENIX Supercomputing Platform'' jointly operated by the Binary Population Synthesis Group and the Stellar Astrophysics Group at Yunnan Observatories, Chinese Academy of Sciences.

\end{acknowledgements} 
        \nocite{*}
        \bibliography{main.bib}
        \bibliographystyle{aa} 

\end{document}